\documentclass[showpacs,showkeys]{revtex4}
\usepackage{amsmath}
\usepackage{amssymb}
\usepackage{graphicx}

\begin{document}

\title{
Chameleon dark matter stars
 }

\author{Vladimir Folomeev,$^{1}$
\footnote{Email: vfolomeev@mail.ru}
Ascar Aringazin,$^{2}$
\footnote{Email: aringazin@gmail.com} and
Vladimir Dzhunushaliev$^{1,2,3}$
\footnote{Email: v.dzhunushaliev@gmail.com}
}
\affiliation{
$^1$Institute of Physicotechnical Problems and Material Science, National Academy of Sciences
of the Kyrgyz Republic, 265 a, Chui Street, Bishkek 720071,  Kyrgyz Republic \\
$^2$Institute for Basic Research,
Eurasian National University,
Astana 010008, Kazakhstan
\\
$^3$ Department of Theoretical and Nuclear Physics, Kazakh National University, Almaty 050040, Kazakhstan
}

\begin{abstract}
We consider static, spherically symmetric
equilibrium configurations consisting of fermionic dark matter
 nonminimally coupled to dark energy in the form of a quintessence scalar field.
With the scalar field coupling function, the form of which is taken to meet cosmological observations,
we estimate the effect of such a nonminimal coupling
on the properties of dark matter compact configurations.
We show that the masses and sizes
of the resulting chameleon dark matter stars are smaller than those of systems with no field present.
\end{abstract}

\pacs{95.36.~+~x, 95.35.~+~d, 04.40.~--~b}
\keywords{Interacting dark matter/dark energy; compact configurations;  mass-radius relation}
\maketitle

\section{Introduction}

During the past one and a half decades,
much observational evidence for the accelerated expansion of the present Universe has appeared~\cite{sahni:2004,Copeland:2006wr}.
Such  acceleration could not be explained only by ordinary matter from which
visible stars and components of galaxies in the Universe are made.
Invisible form of matter called dark energy (DE), which works as a repulsive force against attracting gravity and drives the accelerated expansion, has been invoked. The observational data indicate that about 70\% of the energy density in the present Universe could be assigned to DE.

Another important notion used to explain the evolution of the present Universe is the gravitationally attractive invisible
substance called dark matter (DM). Its contribution
to  the total energy density is estimated to be of order~25\%. DM is clustered on scales of the order of galaxies and clusters of galaxies,
and its existence is so far evident only via its gravitational interaction.

Despite the fact that the true origin of neither DM nor DE is currently known,
various ways have been suggested to model them. The simplest approach is the so-called $\Lambda$CDM model,
where DE is described by Einstein's $\Lambda$ term, and DM is supposed to be
a pressureless fluid
(cold dark matter). However, the well-known cosmological constant problem related to this model
motivates one to look for other ways to describe the accelerated expansion of the Universe.
Perhaps one of the most promising ways to address the origin of DE are theories that include
various fundamental fields~\cite{sahni:2004,Copeland:2006wr}.
Other possible ways discussed in the literature are modified (non-Einstein) gravity theories~\cite{DeFelice:2010aj,Nojiri:2010wj} and models with extra space dimensions \cite{Maartens:2003tw,Dzhunushaliev:2009va}.

Along these lines,
cosmological models with DM and DE interacting with each other not only
gravitationally but also by a direct coupling
\cite{Damour:1990tw,Wetterich:1994bg,Amendola:1999er,Billyard:2000bh,Bartolo:1999sq,Zimdahl:2001ar,
Maccio:2003yk,Farrar:2003uw,Das:2005yj,Amendola:2006qi,Guo:2007zk,Boehmer:2008av,Bean:2008ac}
(for a review, see, e.g., Refs.~\cite{Copeland:2006wr,Tsujikawa:2010sc}) have a number of interesting features,
and may be used in solving
some known problems of  modern cosmology, for example, the so-called coincidence problem.
In turn, the supposed presence in the Universe of DE,  in one form or another,
and DM provides the basis for concluding
that compact systems consisting of dark energy
\cite{de_stars}, of dark matter
\cite{dm_stars,Narain:2006kx}, or of
interacting DE and DM \cite{Brouzakis:2005cj,de_dm_conf}
might also exist.

In the present paper, we study  static configurations supported by  interacting DE
(in the form of a scalar field) and DM (in the form of fermion particles).
Choosing a specific type of interaction, our objective here will be to
identify the effect of this kind of nonminimal coupling
on masses and sizes of resulting compact objects.

The paper is organized as follows. In Sec.~\ref{Lagr_choice} we discuss the choice of the dark matter/dark energy
interaction Lagrangian,  which we use to derive the general set of equations for equilibrium configurations in
Sec.~\ref{gen_eqns}.
Choosing an equation of state of dark matter in the form of an ideal completely degenerate Fermi gas,
in Sec.~\ref{fermi_dark} we write down equations for this particular case and solve them numerically
in Sec.~\ref{fermi_dark_numer}.
 Finally, in Sec.~\ref{conclusion}  our results are summarized.

\section{Equations for equilibrium configurations}
\label{gen_equations_cham_dark}

We study the effect of the presence of direct interaction between a
cosmological  scalar field and dark matter on compact configurations consisting of such dark matter.
Our basic setup is that the dark matter, being embedded in an external, homogeneous cosmological scalar field,
feels its presence not only gravitationally but also through the nonminimal coupling. Obviously,
the characteristics of such mixed configurations will then be determined by the properties both of the dark matter and of the
 scalar field.

\subsection{On the choice of the Lagrangian for the interacting dark matter/dark energy}
\label{Lagr_choice}

At the moment, there exists no fundamental theory that allows to choose
a specific coupling in the dark sector. Therefore, any type of coupling
will necessarily be phenomenological, although some models may appear to be more physically motivated than others.
In modeling DE in the form of a scalar field, one can
meet in the literature different types of coupling  between the field and dark matter.
 One possibility, inspired by
scalar-tensor theories of gravity,  is to consider  an interaction of the form $Q\, T_{(\text{DM})} \varphi_{,\nu}$, where
$T_{(\text{DM})}$ is the trace of the energy-momentum tensor
of dark matter, $\varphi$ is the cosmological scalar field,
 and $Q$ can be a constant \cite{Wetterich:1994bg,Amendola:1999er,Maccio:2003yk} or be field dependent \cite{Billyard:2000bh,Bartolo:1999sq,Amendola:2006qi}.
Another possibility is to modify the continuity equation by including an interaction term of the form 
$\Gamma \rho_{\text{DM}}$, with $\Gamma$ proportional to
the Hubble parameter  \cite{Zimdahl:2001ar,Guo:2007zk,Boehmer:2008av}. In fact, this is a fluid description  of coupled dark energy.

The presence
of a direct coupling between
scalar field and matter (ordinary or dark)
gives rise to a fifth-force interaction  whose magnitude must lie within
the restrictions imposed by local gravity experiments  and observational constraints
from cosmology. In the case of ordinary matter,
this is achieved by suggesting some mechanism
 suppressing the propagation of the fifth force (a screening mechanism).
One possibility is to introduce the so-called chameleon mechanism
\cite{cham_cosm}, which implies that the field
mass depends on the surrounding matter density that effectively results in
decoupling of the field from matter in a high-density background.
Such models are also being  used in describing the present accelerated  expansion of the Universe
(see, however, the recent work of Ref.~\cite{Wang:2012kj}, where it is shown  that the current acceleration
is driven, in fact, by some quintessence field or a cosmological constant, but is not a consequence of gravity modification).

If one initially works within the framework of
general relativity,  the appearance of a direct coupling between scalar field and matter is also possible  when the
mass of matter particles is assumed to be explicitly dependent on the scalar field.
This can be an exponential dependence
(for cosmological implications, see Ref.~\cite{yuk_exp}) or a linear dependence appearing as a consequence
of the presence of the Yukawa coupling. The latter type of interaction  has been repeatedly considered
 in the literature. In particular,  it was used  in studying compact objects in Refs.~\cite{Brouzakis:2005cj,Lee:1986tr,Crawford:2009gx}, 
in describing structure formation  in Ref.~\cite{Nusser:2004qu}, and in modeling  interacting DM and DE
in the papers~\cite{Farrar:2003uw,Bean:2008ac}.

An explicit dependence of the mass of particles on a scalar field implies that by going
to a description of matter in the form of a fluid, its pressure and density  become  functions
depending explicitly on the scalar field (see, e.g., Ref.~\cite{Brouzakis:2005cj}). However, one can  use instead a
phenomenological possibility when a description of the interaction between
scalar field and matter is performed, assuming that  pressure and density are not initially explicit functions of
a scalar field. Such a coupling can be expressed through an interaction  Lagrangian of the form
$
L_{\text{int}}=f(\varphi) L_m,
$
where $L_m$ is the Lagrangian of matter (ordinary or dark) and
the coupling function $f(\varphi)$ characterizes the
interaction between $\varphi$ and the matter.
In this case the
general form of the Lagrangian  can be presented as follows:
 \begin{equation}
\label{lagran_gen}
L=-\frac{c^4}{16\pi G}R+\frac{1}{2}\partial_{\mu}\varphi\partial^{\mu}\varphi -V(\varphi)+f(\varphi) L_m ~.
\end{equation}
The case $f\to 1$ corresponds to the absence of a direct coupling between
matter and scalar field, when the two sources are coupled only via gravity.
Then, for a suitable choice of a potential energy  $V(\varphi)$, such a Lagrangian can describe various  models
of decoupled systems, including quintessence models of dark energy.

In the cosmological context,  the Lagrangian \eqref{lagran_gen} was used  in modeling
the present accelerated  expansion of the Universe
\cite{Beans,Das:2005yj,Koivisto:2005nr,Farajollahi:2010pk,Cannata:2010qd,Folomeev:2012sz},
in describing structure formation \cite{struc_form_f_phi},
and when considering compact configurations
\cite{cham_stars,Folomeev:2012sz}. In doing so, a choice
of the Lagrangian $L_m$ is, in general, not unique.
It can be taken as
$L_m=-\varepsilon$ \cite{Hawking1973}
or  $L_m=p$~\cite{Stanuk2}, where $\varepsilon$ and $p$ are the energy density and the  pressure of an isentropic perfect fluid.
By varying both these  matter Lagrangians with respect to a metric, one obtains the same energy-momentum tensor
of the perfect
fluid in the conventional form. However, it can be shown that, for instance, for static configurations, these Lagrangians,
being substituted  in the general  Lagrangian \eqref{lagran_gen}, will give different equations for an equilibrium
configuration [the Tolman-Oppenheimer-Volkoff (TOV) equations]. In the case of
 $L_m=p$, the TOV equation will have the same form as when $f=1$ \cite{cham_stars,Folomeev:2012sz}.
 On the other hand, the use of
$L_m=-\varepsilon$ leads to the appearance
of an extra term on the right-hand side of the TOV equation
associated with the nonminimal coupling [see Eq.~\eqref{conserv_2_cham_star} below].
Of course, in the limit $f\to 1$, for both choices of $L_m$,
the gravitating system of a  scalar field plus matter
 reduces   to the same uncoupled system in which the matter and the scalar field interact only gravitationally.
But in the general case of $f=f(\varphi)$, these two systems will not be equivalent, giving equilibrium configurations
with different properties.
Apparently, at the present time, it is difficult to make a motivated choice among these Lagrangians
 $L_m$, or any other Lagrangians  used in the literature
(for the other possible Lagrangians and discussion of their choice for nonminimally coupled systems,
see, e.g., Ref.~\cite{Bertolami:2008ab}). For this reason, a description of various systems with nonminimal coupling of the type
 $f(\varphi) L_m$ is made, in effect,
using an {\it ad hoc} choice of  $L_m$.

\subsection{General set of equations}
\label{gen_eqns}

In the present paper, we employ the model described by the general Lagrangian \eqref{lagran_gen} with the dark matter
Lagrangian  $L_{\text{DM}}=-\varepsilon$. Such a choice allows one
to describe naturally compact objects embedded in an external, homogeneous cosmological scalar field
 (see  Sec. \ref{prob_stat} below).
The Lagrangian \eqref{lagran_gen} also
contains  two functions of the scalar field: the potential energy
$V(\varphi)$ and the nonminimal coupling function $f(\varphi)$.
Their form is specified proceeding from some general field-theoretical considerations
in such a way as to be compatible with  current astronomical and cosmological observations.
Note  that since in the model under consideration
the scalar field is coupled only to dark matter, but not to ordinary matter,  it is not necessary to satisfy
constraints from solar-system tests of gravity.

In general,
the presence of interaction between DM and DE may have potentially observable cosmological implications  \cite{Damour:1990tw,Amendola:1999er,Maccio:2003yk,Das:2005yj,Amendola:2006qi,Guo:2007zk,Boehmer:2008av}.
On the other hand,
at astrophysical scales, due to gravitational instabilities, interacting DM and DE may
form compact dense objects composed primarily of dark matter. Clearly, physical properties of such
mixed configurations will depend on the form both of the coupling function and of the
potential energy of the scalar field.

To clarify the question of how the presence of the nonminimal coupling
influences  masses and sizes of the resulting compact gravitating configuration,
 we start from a configuration consisting only of fermionic dark matter, with no scalar field present.
Such objects can support themselves against gravitational contraction
by the degeneracy pressure of fermions obeying Pauli principle, in the same way as in the case of neutron stars and white dwarfs.
After introducing the nonminimal coupling to this system, we will track changes in its properties as they depend on
the parameters appearing in the functions~$f(\varphi)$ and $V(\varphi)$.

We will use Einstein's gravitational equations that allow us to take into account relativistic effects. As a matter source in these equations, we take the energy-momentum tensor, which is
obtained by varying the matter part of the Lagrangian \eqref{lagran_gen} with respect to the metric,
\begin{equation}
\label{emt_cham_star}
T_i^k=f\left[(\varepsilon+p)u_i u^k-\delta_i^k p\right]+ \partial_{i}\varphi\partial^{k}\varphi
-\delta_i^k\left[\frac{1}{2}\,\partial_{\mu}\varphi\partial^{\mu}\varphi-V(\varphi)\right]~,
\end{equation}
where $\varepsilon$ and $p$ are the energy density and the pressure of a dark matter fluid, respectively, and $u^i$ is the four-velocity.

To describe a compact spherically symmetric configuration, we take a static metric of the form
\begin{equation}
\label{metric_sphera}
ds^2=e^{\nu} d(x^0)^2-e^{\lambda}dr^2-r^2d\Omega^2,
\end{equation}
where $\nu$ and $\lambda$ are functions of the radial coordinate $r$, the time coordinate  $x^0=c\, t$,
and $d\Omega^2$ is the metric on the unit two-sphere. Using this metric and the energy-momentum tensor \eqref{emt_cham_star},
we obtain three equations (hereafter we use units where $c=\hbar=1$). First and second are the
$(_0^0)$ and $(_1^1)$ components of the Einstein equations,
\begin{eqnarray}
\label{Einstein-00_cham_star}
&&G_0^0=-e^{-\lambda}\left(\frac{1}{r^2}-\frac{\lambda^\prime}{r}\right)+\frac{1}{r^2}
=8\pi G \left(f \varepsilon +\frac{1}{2}\,e^{-\lambda} \varphi^{\prime 2}+V\right),\\
\label{Einstein-11_cham_star}
&&G_1^1=-e^{-\lambda}\left(\frac{1}{r^2}+\frac{\nu^\prime}{r}\right)+\frac{1}{r^2}
=8\pi G \left(-f p -\frac{1}{2}\,e^{-\lambda} \varphi^{\prime 2}+V\right),
\end{eqnarray}
and the third one follows
from the
law of conservation of energy and momentum,
$T^k_{i;k}=0$. Taking the $i=1$ component of this equation gives
\begin{equation}
\label{conserv_2_cham_star}
\frac{d p}{d r}=-(\varepsilon+p)\left(\frac{1}{2}\nu^\prime+\frac{1}{f}\frac{d f}{d\varphi}\varphi^\prime\right).
\end{equation}
In the above equations, the prime denotes differentiation with respect to $r$.
As has already been discussed at the end of Sec.~\ref{Lagr_choice},
because of the choice of the dark matter Lagrangian  in the form
$L_{\text{DM}}=-\varepsilon$,
 Eq.~\eqref{conserv_2_cham_star} becomes different
from the usual equation of hydrostatic equilibrium:
in the right-hand side an extra term  (the ``fifth'' force) associated with the coupling function $f$ appears.
 Another possible choice, $L_m=p$,
 used in Refs.
\cite{cham_stars,Folomeev:2012sz} implies that
the term containing $f$ in the right-hand side of Eq.~\eqref{conserv_2_cham_star} is absent.

Equations \eqref{Einstein-00_cham_star}-\eqref{conserv_2_cham_star} must be supplemented by an equation
for the scalar field that follows from the Lagrangian \eqref{lagran_gen}:
 \begin{equation}
\label{sf_eq_gen}
\frac{1}{\sqrt{-g}}\frac{\partial}{\partial x^i}\left[\sqrt{-g}g^{ik}\frac{\partial \varphi}{\partial x^k}\right]=
-\left(\frac{d V}{d \varphi}+\varepsilon \frac{d f}{d \varphi}\right).
\end{equation}

Thus, we have five unknown functions:
$\nu, \lambda, \varphi, \varepsilon$, and $p$.
Keeping in mind that $\varepsilon$ and $p$ are related by an  equation of state,
there remain only four unknown functions.
To determine these functions, we have four equations: the
two Einstein equations \eqref{Einstein-00_cham_star} and \eqref{Einstein-11_cham_star},
the scalar-field equation \eqref{sf_eq_gen},
and the equation of hydrostatic equilibrium \eqref{conserv_2_cham_star} [the modified Tolman-Oppenheimer-Volkoff equation].

\subsection{The case of fermionic dark matter}
\label{fermi_dark}

In order to perform calculations, we should choose some equation of state for dark matter.
To elucidate how the interaction
between DM and DE affects the characteristics of compact configurations,
we will restrict ourselves to some simplest form of an equation of state.
Namely, we assume that dark matter is an
ideal completely degenerate Fermi gas at zero temperature. Its equation of state
is obtained using usual expressions
for the energy density and pressure~\cite{Narain:2006kx,Shapiro:1983}:
\begin{eqnarray}
\label{eos_enrg_dens}
&&\varepsilon=\frac{1}{\pi^2}\int_0^{k_F}k^2\sqrt{m_f^2+k^2}dk=
\frac{m_f^4}{8\pi^2}\left[z\sqrt{1+z^2}(1+2 z^2)-\sinh^{-1}(z)\right]\equiv m_f^4 \tilde{\varepsilon}
,\\
\label{eos_p}
&&p=\frac{1}{3\pi^2}\int_0^{k_F}\frac{k^4}{\sqrt{m_f^2+k^2}}dk=
\frac{m_f^4}{24\pi^2}\left[z\sqrt{1+z^2}(2 z^2-3)+3 \sinh^{-1}(z)\right]\equiv m_f^4 \tilde{p}.
\end{eqnarray}
Here $m_f$ is  the fermion mass, $k_F$ is  the Fermi momentum, $z=k_F/m_f$  is the relativity parameter, and
$\tilde{\varepsilon}$ and $\tilde{p}$ are
the dimensionless energy density and pressure expressed in units of $m_f^4$.

In two limiting cases, this equation of state can be represented in simple power-law forms:
(i) in the nonrelativistic case, $z\ll 1$, we get the polytropic law, $\tilde{p} \propto \tilde{\varepsilon}^{5/3}$, and
(ii) in the ultrarelativistic case, $z\gg 1$, we have $\tilde{p} = \tilde{\varepsilon}/3$.

Next, introduce dimensionless variables for the current radius $r$  and mass $M(r)$
of the configuration under consideration~\cite{Narain:2006kx},
\begin{equation}
\label{dimless_xi_v}
\tilde{M}=\frac{M}{M_L}, \quad \tilde{r}=\frac{r}{R_L} \quad
\text{with} \quad M_L=\frac{M_{\text{Pl}}^3}{m_f^2}, \quad R_L=\frac{M_{\text{Pl}}}{m_f^2},
\end{equation}
where $M_{\text{Pl}}$ is the Planck mass. The parameters
$M_L$ and $R_L$ are the characteristic mass and radius
obtained by Landau in considering compact configurations consisting of an ultrarelativistic degenerate Fermi gas
within the framework of Newtonian gravity. The expressions for $M_L$ and $R_L$ presented in
Eq.~\eqref{dimless_xi_v} can be obtained as follows (for more details, see Ref.~\cite{Narain:2006kx}).
If one considers a particle with mass $m_f$ located on the surface of a star, its energy is equal to the sum
of the gravitational energy $-G {\cal M} m_f/{\cal R}$ and of the Fermi energy, which is proportional to
$N^{1/3}/{\cal R}$. In these expressions
$N$ is  the total number of fermions in a star, and ${\cal M}\equiv m_f N$ is the total mass of the star
whose radius is ${\cal R}$.
As a good estimate of the maximum possible number of fermions in a star $N_{\text{max}}$, one can consider
the limiting case when the energy of the particle is equal to zero, i.e., the case when
the degeneracy pressure of the fermionic fluid is balanced by the gravitational
attraction.
As a result, this yields
$N_{\text{max}}\sim \left(M_{\text{Pl}}/m_f\right)^3$.
Correspondingly, the maximum mass of the star will then be ${\cal M}_{\text{max}}\sim M_{\text{Pl}}^3/m_f^2$,
which may be regarded as a characteristic mass for such fermionic configurations. The minimum value of the radius,
corresponding to this maximum mass, is obtained by assuming that
the kinetic energy of the fermion becomes
equal to its mass on the surface of a star, i.e., $k_F\approx m_f$.
As a result, one can find the minimum radius
${\cal R}_{\text{min}}\sim M_{\text{Pl}}/m_f^2$, and this corresponds to the
characteristic radius $R_L$ from Eq.~\eqref{dimless_xi_v}.

Now, introducing the dimensionless scalar field, $\phi=\varphi/M_{\text{Pl}}$,
and rewriting the metric function  $\lambda$ in terms of the Schwarzschild mass parameter $M$ as
$e^{-\lambda}=1-2 G M(r)/r$, we represent Eqs.~\eqref{Einstein-00_cham_star}, \eqref{conserv_2_cham_star}, and \eqref{sf_eq_gen}
in the following dimensionless form:
\begin{eqnarray}
\label{TOV_dmls}
&&\frac{d p}{d r}=
-(\varepsilon+p)\left\{\frac{M-4\pi r^3\left[-f p-1/2 (1-2  M/r)\phi^{\prime 2}+V\right]}{r(r-2 M)}
+\frac{1}{f}\frac{d f}{d\phi}\phi^\prime\right\}
,\\
\label{G00_dmls}
&&\frac{d M}{d r}=4\pi r^2 \left[f \varepsilon +\frac{1}{2} \left(1-\frac{2  M}{r}\right)\phi^{\prime 2}+V\right]
,\\
&&\phi^{\prime\prime}+\frac{1}{2}\left\{
\frac{3}{r}+\frac{1-8\pi r^2\left[-f p-1/2 (1-2  M/r)\phi^{\prime 2}+V\right]-
2(M^\prime-M/r)}{r-2 M}\right\}\phi^\prime \nonumber\\
&&=\frac{1}{1-2  M/r}
\left(\frac{d V}{d\phi}+\varepsilon\frac{d f}{d\phi}\right).
\label{sf_eq_dmls}
\end{eqnarray}
For convenience, we hereafter drop the tilde.
Equation \eqref{TOV_dmls} was obtained from  combining Eqs.~\eqref{conserv_2_cham_star} and \eqref{Einstein-11_cham_star},
and Eq.~\eqref{G00_dmls} follows immediately from \eqref{Einstein-00_cham_star}.
Thus, to describe the static configuration under consideration we have obtained three equations \eqref{TOV_dmls}-\eqref{sf_eq_dmls}
which must be supplemented by appropriate
scalar field functions $f(\phi)$ and $V(\phi)$.

\section{Numerical results}
\label{fermi_dark_numer}

\subsection{Statement of the problem}
\label{prob_stat}

As an example, consider here a particular choice of the  functions $f(\phi)$ and $V(\phi)$.
We suppose that our spherically symmetric configuration
is taken to be embedded in an external, homogeneously distributed cosmological scalar field $\phi_0$.
For the scalar  field, we will look for such solutions  that start from some central value
 $\phi_c$ and tend, at large distances,  to
the background  value  $\phi_0$. Then it seems reasonable to choose such function
 $f(\phi)$ which tends, at large  $r$,  to 1. Such behavior of
 $f(\phi)$ implies that the nonminimal coupling works only at relatively high densities of dark matter,
 which are typical for the inner regions of a star. On the other hand,
at relatively low densities of dark matter, which are typical for  interstellar and intergalactic space, the
nonminimal coupling switches off. In turn, the  solution for the scalar field far from the edge
of the configuration smoothly matches the cosmological background value $\phi_0$ (for details, see below).
Then, if one chooses the potential energy  $V(\phi)$, for instance, in the form of a quintessence potential,
the configuration under investigation can be considered as embedded in the Universe described by
quintessence models.

\subsection{An explicit example}

The choice of the scalar functions $f(\phi)$ and $V(\phi)$ is strictly model dependent and is usually done proceeding
from some reasonable motivations. For example, in Ref.~\cite{Beans} the coupling function $f(\phi)$  is taken in
a power-law form which is a consequence of  a conformal transformation from the string frame into the Einstein
frame, and $V(\phi)$ -- in the form of an exponential quintessence potential.
In Ref.~\cite{Farajollahi:2010pk} $f(\varphi)$  and  $V(\varphi)$
are taken to be  exponential and power functions, and their parameters are chosen in such a way as to satisfy the current observational data.
Another approach has been applied in
Refs.~\cite{Cannata:2010qd,Folomeev:2012sz}, where the authors have initially selected some particular  form of cosmological evolution and  found the functions  $f(\varphi)$  and  $V(\varphi)$ corresponding to such evolution.

Here we choose the functions $f(\phi)$ and $V(\phi)$ which were used in
 Ref. \cite{Das:2005yj} to model the  evolution of the present Universe. In our dimensionless variables, they can be
 rewritten as follows:
\begin{equation}
\label{scal_func_dmls}
f(\phi)=\exp\left[\beta (\phi-\phi_0)\right], \quad V(\phi)=\delta (1/\phi)^\alpha \quad \text{with} \quad \delta=\left(M_\varphi/m_f\right)^4,
\end{equation}
where  $\alpha$, $\beta$, and $\phi_0$ are positive quantities,
with $\beta \sim {\cal O}(1)$,
and the mass scale is tuned to $M_\varphi\sim 10^{-3}~\text{eV}$ in order for acceleration to occur at the present
epoch.
In this model, $\phi_0$ corresponds to the current value of the cosmological scalar field.
Note that the potential $V(\varphi)$ is an example of a tracker potential in quintessence models \cite{Ratra:1987rm}.
Moreover, the above functions \eqref{scal_func_dmls} are similar to those
used within the framework of
chameleon cosmology \cite{cham_cosm},
where the direct coupling between a cosmological scalar field and ordinary (baryon) matter takes place.

We solve the system of equations \eqref{TOV_dmls}-\eqref{sf_eq_dmls} numerically for given $\alpha$, $\beta$, and $\delta$ subject to the following boundary conditions in the vicinity
of the center of the configuration $r=0$,
\begin{equation}
\label{bound_all}
\varepsilon \simeq \varepsilon_c+\frac{\varepsilon_2}{2}r^2, \quad M \simeq M_3 r^3, \quad
\phi \simeq \phi_c+\frac{\phi_2}{2}r^2,
\end{equation}
where $\phi_c$ and $\varepsilon_c$ denote the central values of
$\phi$ and the energy density $\varepsilon$ at $r=0$. The coefficients $\varepsilon_2$, $M_3$, and $\phi_2$ can be found
from Eqs.~\eqref{TOV_dmls}-\eqref{sf_eq_dmls}. In solving these equations, we will use
the dimensionless equation of state which is given parametrically
by Eqs.~\eqref{eos_enrg_dens} and \eqref{eos_p} in terms of the variable $z$.

We start the numerical procedure near the origin $r \approx 0$ and proceed to the point
$r = r_b$,  where the pressure is zero. We refer to the obtained solutions as internal solutions.
The mass contained inside the sphere of the radius $r_b$ will be treated as the mass of the dark matter star.

Since the spherically symmetric configuration under consideration
is supposed to be embedded in an external, homogeneously distributed cosmological scalar field $\phi_0$, then,
to provide the smoothness of solutions along the radius,
we require the internal solutions to sew solutions obtained for the region $r > r_b$ characterized by a nonzero scalar field energy density.
Thus, for $r > r_b$ we proceed numerical solutions
of Eqs.~\eqref{G00_dmls} and \eqref{sf_eq_dmls} retaining
only the gravitational and scalar fields while the dark matter fluid is taken to be zero.
At large distances, the scalar field reaches its asymptotic value $\phi_0$ taken from
Ref.~\cite{Das:2005yj}, which can be rewritten
in our dimensionless variables as follows:
\begin{equation}
\label{phi_0}
\phi_0\approx \frac{\alpha}{\sqrt{8\pi}\beta}\frac{\Omega_{\text{DE}}^{(0)}}{\Omega_{\text{DM}}^{(0)}}.
\end{equation}
Here $\Omega_{\text{DE}}^{(0)}$ and $\Omega_{\text{DM}}^{(0)}$
denote the dimensionless densities of dark energy and dark matter, respectively,
measured in units of the current critical density. The factor $1/\sqrt{8\pi}$ appears here since
we take
$M_{\text{Pl}}=1/\sqrt{G}$ instead of the reduced Planck mass,
$M_{\text{Pl}}=1/\sqrt{8\pi G}$, used in Ref.~\cite{Das:2005yj}.
This value of $\phi_0$ provides  the current averaged cosmological density in the Universe
$\backsimeq 10^{-47} \text{GeV}^{4}$.

Formally, at the points where the scalar field becomes equal to $\phi_0$, there is a nonzero gradient of the field.
However, since for the values of the parameters $\alpha$, $\beta$, and $\delta$ being used here (see below)
the field goes to $\phi_0$ only at $r\gg r_b$, the magnitude of the gradient part of the energy density $T_0^0$
is always much smaller than the potential energy $V(\phi)$. This allows us to consider the field in this region,
to a good approximation, as homogeneous.

Using the above procedure, we define the mass of the configuration under consideration as the mass of all matter
(dark matter plus dark energy) contained inside the sphere of the radius  $r = r_b$.
In Ref.~\cite{cham_stars} configurations consisting of ordinary (baryon) matter
nonminimally coupled to a   scalar field
 were called ``chameleon stars.'' The reason is that  the presence of such an
interaction results in a substantial change of the inner structure of a star.
 By analogy, we can call the configurations being considered here
 chameleon dark matter stars and denote their mass as $M_{\text{CDMS}}\equiv M(r_b)$.

Another physically relevant parameter is given
by the coordinate $\bar{r}$
associated with the proper radius of the star.
This quantity is defined so as to be invariant with respect to spatial coordinate transformations
preserving the spherical symmetry, and it can be presented
 as follows:
\begin{equation}
\label{xi_observ}
\bar{r}=\int_{0}^{r} e^{\lambda/2} d r^\prime=\int_{0}^{r} \left[1-\frac{2 M(r^\prime)}{r^\prime}\right]^{-1/2} d r^\prime.
\end{equation}
Then, in dimensional variables, the proper radius is
$R_{\text{prop}}=\bar{r}_{b} R_L$.
Since $R_L$
is equal to half the Schwarzschild radius, then we require $\bar{r}_{b}>2$ to avoid black hole configurations.

\begin{figure}[t]
\centering
  \includegraphics[height=10cm]{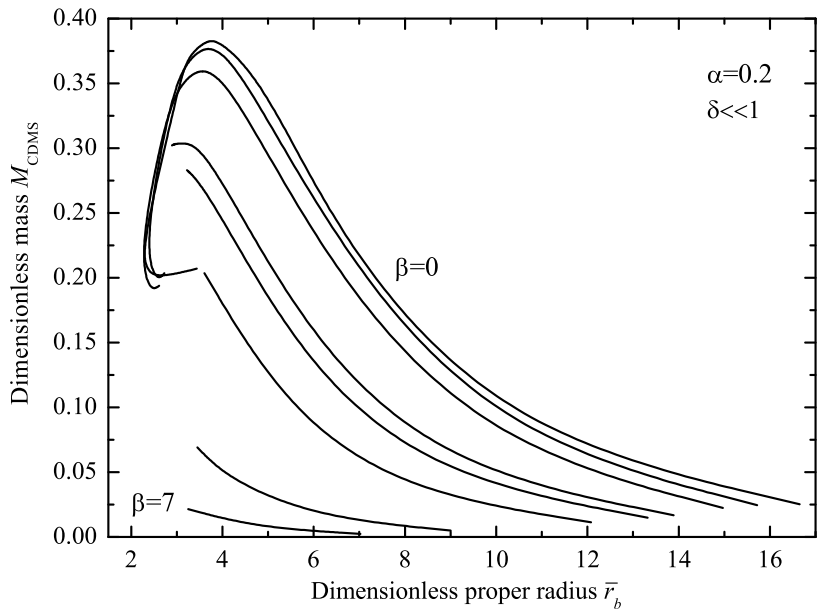}
\vspace{-1.cm}
\caption{The mass-radius relation for configurations with equation of state
 given by Eqs.~\eqref{eos_enrg_dens} and \eqref{eos_p} and functions $f, V$ from
\eqref{scal_func_dmls}. The parameter $\beta=0, 0.5, 1, 2, 2.3, 3, 5, 7$,
from top to bottom. The curve labeled by $\beta=0$ corresponds to configurations without a scalar field.
}
\label{fig_M_R}
\end{figure}

We turn now to a consideration of solutions obtained according to the above procedure.
To perform numerical calculations, it is necessary to choose the values of the
parameters $\alpha$, $\beta$, and $\delta$. Here we fix
 $\alpha=0.2$, which has been used in the paper
\cite{Das:2005yj} in describing the evolution of the Universe.
The value of the parameter $\delta$
appearing in
Eq.~\eqref{scal_func_dmls} depends both
on the mass scale parameter $M_\varphi$ (in Ref.~\cite{Das:2005yj} it was taken as $M_\varphi\sim 10^{-3}\,\text{eV}$)
and on the fermion mass $m_f$.
Since at the moment  it is not definitely known which type of fermion  makes up  gravitationally bound
clumps of dark matter,
various fermion particles are considered in the literature.
This could
be both  superlight gravitinos with a mass of the order of
$10^{-2}~\text{eV}$ and superheavy  WIMPs with a TeV mass scale \cite{Bertone:2004pz}.
Following Ref.~\cite{Narain:2006kx}, we assume for definiteness that
$m_f$ lies in the range $1~\text{eV} \lesssim m_f \lesssim 10^{2}~\text{GeV}$. Then the value of the parameter
$\delta$ is always considerably smaller than unity, and, as the numerical calculations indicate,
the influence of the potential  $V(\phi)$
from
\eqref{scal_func_dmls} on the solutions is negligibly small
compared with other terms; i.e., the scalar field can be treated as having no potential energy
on the scales associated with the fermion mass scale $m_f$.

Thus, we have only one free parameter  $\beta$ whose value is assumed to be
${\cal O}(1)$ \cite{Das:2005yj,cham_cosm},
corresponding to gravitational strength coupling, but can be even much larger \cite{Mota:2006ed}
(see also Ref.~\cite{struc_form_f_phi} where the coupling constant is taken to be negative).
Since its explicit value does not follow
from any first principles,  we will vary its magnitude slightly, keeping track of changes in
characteristics of compact configurations being considered~here.

\begin{figure}[t]
\begin{minipage}[t]{.5\linewidth}
  \begin{center}
  \includegraphics[width=9cm]{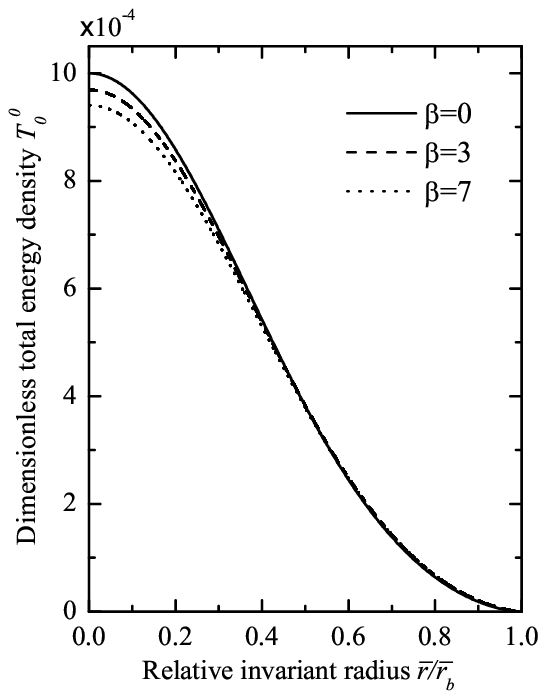}
  \end{center}
\end{minipage}\hfill
\begin{minipage}[t]{.5\linewidth}
  \begin{center}
  \includegraphics[width=9cm]{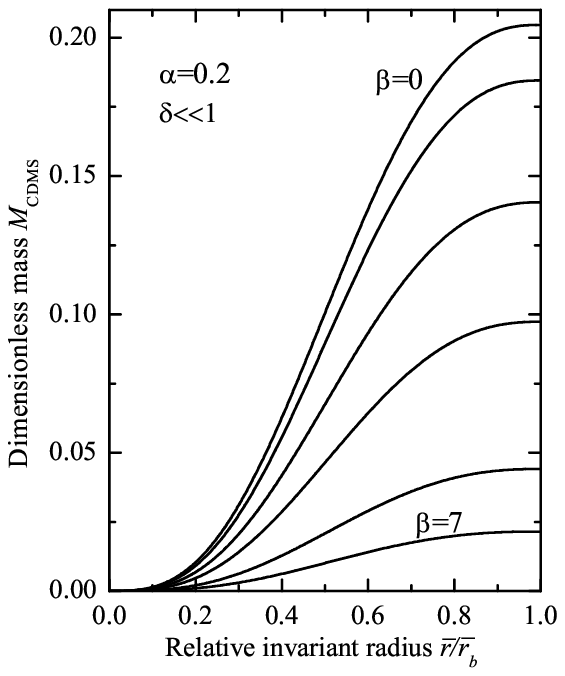}
  \end{center}
\end{minipage}\hfill
\vspace{-1.cm}
  \caption{Typical distributions of the dimensionless total energy density $T_0^0$
from Eq.~\eqref{Einstein-00_cham_star} (left panel) and of the current masses $M_{\text{CDMS}}=M(\bar{r})$ (right panel)
as functions of the  relative invariant radius $\bar{r}/\bar{r}_b$ for different $\beta$. For both panels,
the central dark matter energy density is taken as $\varepsilon_c=10^{-3}$. In the right panel,
the parameter $\beta$ runs the values $0,  1, 2, 3, 5, 7$,
from top to bottom.
  }
 \label{energ_mass}
\end{figure}

The results of numerical calculations of Eqs.~\eqref{TOV_dmls}-\eqref{sf_eq_dmls}
with the boundary conditions \eqref{bound_all} are presented in Figs.~\ref{fig_M_R} and \ref{energ_mass}.
Figure~\ref{fig_M_R} shows the mass-radius relation for different values of the parameter $\beta$.
In plotting the curves, we have used
the values of the dimensionless central density of dark matter
$\varepsilon_c$ from Eq.~\eqref{bound_all} lying in the range  $10^{-5}\leq\varepsilon_c \leq 10$.

In obtaining the solutions, we started from such central values
$\phi_c$ at which the scalar field asymptotically, when $r\gg r_b$, went to
the background value $\phi_0$ from Eq.~\eqref{phi_0}.
The numerical calculations indicate that
 $\phi_c$ must be always less than that  given by Eq.~\eqref{phi_0}. Since
$\phi_0$ depends on $\beta$, a situation
may occur when for some $\beta$ and
$\varepsilon_c$ it is necessary to set $\phi_c<0$.
It is therefore inevitable in this case that $\phi(r)$ would cross zero
at some point, giving a singularity in the potential
$V(\phi)$, that is physically unacceptable. That is why, in plotting the curves in Fig.~\ref{fig_M_R},
we restrict ourselves to only such values of
 $\varepsilon_c$ which provide the positivity of $\phi(r)$ everywhere along the radius.
 This implies that not all
mass-radius curves have a  maximum.
There exists some critical value of
$\beta \approx 2.3$ which still allows the curve to approach the maximum at some
 $\varepsilon_c$. At larger $\beta$, the maximum is absent.

In Fig.~\ref{fig_M_R}, the curve labeled $\beta=0$ corresponds to the case studied in Ref.~\cite{Narain:2006kx}
with no scalar field  present.
The points of the curve to the right of the  maximum correspond to stable configurations:
with increase of the mass of a star its radius decreases.
The points of the curve to the left of the  maximum refer to unstable configurations.
The curves for configurations with the scalar field demonstrate
the similar behavior
for systems with
$\beta \lesssim 2.3$ when the maximum is still present.
At bigger values of $\beta$, the absence of  maxima means that
all configurations are stable.

By comparing the stable configurations with and without the scalar field,
one can identify the changes brought about by the presence of the nonminimal coupling. Namely,
with the increase of $\beta$ that corresponds to a stronger interaction
between dark matter and dark energy,
both the masses (at a fixed star radius) and the sizes (at a fixed star mass)
of the systems with the field become smaller, as compared to the configurations without a scalar  field.
Moreover, the differences between masses and sizes in these two cases can be as high as the order of magnitude, at large values of $\beta$
(cf. Fig.~\ref{energ_mass}).

These results agree with those we get when considering the radial distribution of the matter.
The results of  calculations with some fixed central value of the
dark matter energy density $\varepsilon_c$ are shown in Fig.~\ref{energ_mass}.
As one can see from the plots presented in the left
panel, configurations with the scalar field are characterized by a lower concentration of the matter
towards the center
(compare these results with those of Ref.~\cite{struc_form_f_phi} where the influence
of the nonminimal coupling between scalar field and dark matter on the structure
of galactic haloes was studied).
Also, taking into account the fact
that the systems with the scalar field are more compact as compared to the field-free configurations having the same
$\varepsilon_c$, we eventually obtain configurations whose masses
are less than that  of the system without a scalar field,
as demonstrated in the right panel of Fig.~\ref{energ_mass}.

Let us now derive an approximate formula
for the dependence of the maximum mass of the stars
on the parameter $\beta$. To do this, first we convert the dimensionless variables \eqref{dimless_xi_v} to the physical ones.
We recall also that not all mass-radius curves have a  maximum. So, for the configurations with $\beta \gtrsim 2.3$,
we take the very left point of the mass-radius curve as corresponding to the maximum mass.
Following Ref.~\cite{Narain:2006kx}, we take the characteristic fermion mass scale as
$m_f=1~\text{GeV}$, for which the characteristic Landau mass $M_L=1.632 M_\odot$. Then the
corresponding approximate expression for a maximum mass, $M_{\text{max}}=M_{\text{max}}(\beta)$,
can be presented in the form
\begin{equation}
\label{M_max_beta}
M_{\text{max}}\approx 0.627 M_\odot \left(\frac{1~\text{GeV}}{m_f}\right)^2 \, e^{-0.1\, \beta^{5/3}}.
\end{equation}

This expression gives masses of the order of a stellar mass for the fermion mass $m_f \sim 1~\text{GeV}$.
In the two extreme cases we have the following: (i) for superlight particles of mass $m_f \sim 1~\text{eV}$,
we get very heavy and large configurations of mass
 $M_{\text{max}} \sim 10^{17}-10^{18} M_\odot$ and the radius of the order of a typical
galaxy cluster size, $R\sim 10^{24}~\text{cm}$ and (ii)~for superheavy fermions of mass $m_f \sim 100~\text{GeV}$,
we get light and small objects of mass $M_{\text{max}} \sim 10^{-3}-10^{-4} M_\odot$
and  $R\sim 10^{2}~\text{cm}$.
The characteristics of realistic objects of this type, if they exist at all, will evidently lie somewhere between these
limiting values.

\section{Conclusion}
\label{conclusion}

We have studied the model describing compact gravitating configurations consisting of interacting dark matter and dark
energy. The latter is modeled by the  scalar field
$\varphi$,  whose nonminimal coupling to dark matter is
described by the function $f(\varphi)$ in the Lagrangian \eqref{lagran_gen}.
In general, the form of this function does not follow from first principles, and it must be chosen from some
additional considerations. As an  example,
here we take this function in the form of Eq.~\eqref{scal_func_dmls}, which was
used in Ref.~\cite{Das:2005yj} to describe the present accelerated expansion of the Universe. For such a choice,
the particular case of a compact configuration with dark matter represented as an ideal completely degenerate Fermi gas has been studied.
In order to elucidate the role of the scalar field,
we made a comparison of dark matter configurations supported only by the Fermi gas with configurations containing the nonminimal
coupling.
In this case we showed the following:
\vspace{-0.2cm}
\begin{enumerate}
\itemsep=-0.2pt
\item[(1)] 
There exist regular static solutions found numerically by solving the coupled Einstein-matter
equations subject to a set of appropriate boundary conditions.
The obtained solutions describe compact
mixed dark matter/dark energy configurations, whose  main mass  is concentrated inside the
radius corresponding to the edge of the dark matter, $r=r_b$, where the pressure and density of the dark matter vanish.
\item[(2)] Assuming that the effects of the nonminimal coupling are only essential  at relatively high densities of dark matter,
we sought such solutions for the scalar field that were started
from some central value $\phi_c$ and went to the cosmological background value $\phi_0$
at $r\gg r_b$. This provides that the nonminimal coupling function $f(\phi)$ from Eq.~\eqref{scal_func_dmls} tends to 1 asymptotically,
and such configurations can be considered as embedded
in an external, homogeneous cosmological scalar field
whose density is close to the critical one $\backsimeq 10^{-47} \text{GeV}^{4}$.
In other words,
the configurations under consideration may be thought of as embedded in the Universe described
by the quintessence  Lagrangian \eqref{lagran_gen} with $f=1$.
One might then expect that the quintessence potential  $V(\phi)$ could be any other   function
compatible with  observations,
and not just that given by
Eq.~\eqref{scal_func_dmls} (in this connection see also item (4) below).
\item[(3)] 
The  parameter $\beta $ appearing in the coupling function $f(\phi)$ and determining
the strength of coupling
largely influences  masses and sizes of the objects under consideration.
In particular,
as  is seen from Fig.~\ref{fig_M_R},
while $\beta$ increases,
both the masses (at a fixed star radius) and the sizes (at a fixed star mass)
of stars with the nonminimal
coupling tend to become smaller than those of configurations without a scalar field.
\item[(4)]  On the scales under consideration, the potential energy $V(\phi)$
has no substantial influence on the characteristics of the objects under investigation.  This allows us to suppose
that the use of another quintessence potential will give rise to configurations having similar physical parameters.
\end{enumerate}
\vspace{-0.2cm}

Using the analogy with  chameleon gravity \cite{cham_cosm},
where the scalar field is strongly coupled to  matter and has a mass that
depends on the density of surrounding matter,
here we call the configurations under consideration   chameleon dark matter stars since their mass
depends considerably on the properties of the surrounding scalar field nonminimally coupled to the dark matter.
This name implies that in modeling
dark matter by some equation of state relating the initially
$\phi$-independent
energy density and pressure, the properties of resulting compact objects
will depend substantially on the form of the coupling.

With the choice of $f(\phi)$ in the form of Eq.~\eqref{scal_func_dmls}
and the conditions given in Sec.~\ref{prob_stat}, solutions describing
equilibrium configurations can exist only when  $f(\phi)<1$
along the radius of the configuration.  This results in a smaller
concentration of the matter
along the radius for any $\beta > 0$ (see the left panel of Fig.~\ref{energ_mass}),
and correspondingly leads to less massive configurations,
as compared  with the case of a star without a scalar field
(see the right panel of Fig.~\ref{energ_mass}). It is evident that  another more or less reasonable choice of
 $f(\phi)$, providing the existence of regular solutions
with $f(\phi)>1$ along the radius,
might give rise to configurations having a greater concentration of matter, and correspondingly  a greater mass
(cf. the configurations of Ref.~\cite{cham_stars}, where systems
consisting of the scalar field nonminimally coupled to ordinary matter have been considered).
In that case,  one might naively expect that masses and sizes of resulting configurations would still be determined
basically by the form of  $f(\phi)$, but not by  the quintessence potential  $V(\phi)$.
This  must be investigated in further studies.

\section*{Acknowledgements}

V.D. and V.F. gratefully acknowledge support provided by the Volkswagen Foundation.
This work was partially supported by Grants No.~378 and No.~1626/GF3 in fundamental research in natural sciences
by the Ministry of Education and Science of Kazakhstan.

\end{document}